\documentclass[a4paper,11pt]{article}
\pdfoutput=1
\usepackage[utf8]{inputenc}
\usepackage[T1]{fontenc}
\usepackage{jheppub}
\usepackage{natbib}

\usepackage[usenames,dvipsnames,svgnames,table,x11names]{xcolor}
\usepackage{lmodern,amsmath,multirow,cancel}
\usepackage{booktabs}
\usepackage{xstring}
\usepackage{ifthen}

\usepackage{ascmac,braket,bm,mathrsfs,amsthm,amsfonts}
\usepackage{hyperref}
\usepackage{graphicx}
\usepackage{subcaption} % Allow for subfigures
\usepackage{comment}
\usepackage[normalem]{ulem}

\usepackage{titlesec}
\newcommand*{\justifyheading}{\raggedright}
\titleformat{\chapter}[display]
  {\normalfont\huge\bfseries\justifyheading}{\chaptertitlename\ \thechapter}
  {20pt}{\Huge}
\titleformat{\section}
  {\normalfont\Large\bfseries\justifyheading}{\thesection}{1em}{}
\titleformat{\subsection}
  {\normalfont\large\bfseries\justifyheading}{\thesubsection}{1em}{}
\titleformat{\subsubsection}
  {\normalfont\bfseries\justifyheading}{\thesubsubsection}{1em}{}
\usepackage[english]{babel}
\usepackage[dvipsnames]{xcolor}
\numberwithin{equation}{section}

\renewcommand{\star}{\ast}

\usepackage{slashed}

\renewcommand\Re{\mathop{\mathrm{Re}}}

\usepackage{environ}
\newcommand{\rtext}[1]{\textcolor{red}{#1}}
\newcommand{\ocom}[1]{\textcolor{purple}{[{\rm Osamura: #1}]}}

\newcommand{\sla}[1]{\slashed{#1}}

\newcommand{\opn}[1]{\operatorname{#1}}
\newcommand{\re}[1]{\opn{Re} \left[ #1 \right]}
\newcommand{\im}[1]{\opn{Im} \left[ #1 \right]}
\NewEnviron{eqsp}{%
\begin{equation}\begin{split}
  \BODY
\end{split}\end{equation}
}
\renewcommand{\proofname}{$\because$}

\usepackage{adjustbox}
\usepackage{tabularx}
\newcolumntype{Y}{>{\centering\arraybackslash}X} %for tabularx
\usepackage{booktabs} %for \toprule
\usepackage{colortbl} %for \rowcolors
\def\beq#1\eeq{\begin{align}#1\end{align}}
\RequirePackage{xspace}
\def\Bbar    {\kern 0.18em\overline{\kern -0.18em B}{}\xspace}

\newcommand{\eq}[1]{Eq.~(\ref{#1})}

\newcommand{\tk}[1]{\textcolor{orange}{#1}}

\preprint{CHIBA-EP-267, IPMU24-0005}

\title{
Two-loop corrections to QCD \boldmath{$\theta$} angle from evanescent operator in the BMHV scheme
} 

\author[a]{Tatsuya Banno,}
\author[a,b,c]{Junji Hisano,}
\author[d,b]{Teppei Kitahara,}
\author[a]{Kiyoto Ogawa,}
\author[a]{and\\ Naohiro Osamura}

\affiliation[a]{
  Department of Physics, Nagoya University, Furo-cho Chikusa-ku, Nagoya 464-8602 Japan
}
\affiliation[b]{
  Kobayashi-Maskawa Institute for the Origin of Particles and the
  Universe, Nagoya University,
  Furo-cho Chikusa-ku, Nagoya 464-8602 Japan
}
\affiliation[c]{
  Kavli IPMU (WPI), UTIAS, The University of Tokyo, Kashiwa 277-8584, Japan
}

\affiliation[d]{
Department of Physics, Graduate School of Science,
Chiba University, Chiba 263-8522, Japan}

\emailAdd{banno.tatsuya.p8@s.mail.nagoya-u.ac.jp}
\emailAdd{hisano@eken.phys.nagoya-u.ac.jp}
\emailAdd{kitahara@chiba-u.jp}
\emailAdd{ogawa.kiyoto.f8@s.mail.nagoya-u.ac.jp}
\emailAdd{osamura.naohiro.j2@s.mail.nagoya-u.ac.jp}

\abstract{
We study the renormalization of the QCD $\theta$ angle at the two-loop level focusing on divergent and finite $CP$-violating contributions from evanescent operators, using dimensional regularization with the BMHV scheme.
When one considers the Lagrangian in $d$-dimensional space-time instead of four dimensions in dimensional regularization, evanescent operators that break the chiral symmetry are induced.
Consequently, $T$-odd and $P$-odd fermion loops in the BMHV scheme generate evanescent contributions to the QCD $\theta$ angle. 
We carefully classify the evanescent contributions into two types: 
one originating from the evanescent operators at the one-loop level and the other directly produced by two-loop calculations. 
We show that renormalization of the parameters in the BMHV scheme keeps removing those unphysical contributions to 
the  QCD $\theta$ angle at any scale at the two-loop level.
We also discuss how the rephasing invariance of the radiative correction to the QCD $\theta$ angle is realized in the BMHV scheme.
}


\begin{document}
\sloppy %https://tex.stackexchange.com/questions/9107/how-can-i-make-my-text-never-go-over-the-right-margin-by-always-hyphenating-or-b

% get rid of JHEP header
\makeatletter\renewcommand{\@fpheader}{\ }\makeatother

\maketitle

\renewcommand{\thefootnote}{\#\arabic{footnote}}
\setcounter{footnote}{0}

%=======================================================
%        memo
%=======================================================
%\section*{memo}
%\begin{itemize}
%\end{itemize}

%=======================================================
%        INTRODUCTION
%=======================================================
\section{Introduction}
\label{sec:introduction}

The strong $CP$ problem is a fundamental issue in the Standard Model (SM) of particle physics. It is associated with operators in the QCD Lagrangian that violate both parity ($P$) and time-reversal ($T$) symmetries:
\begin{align}
    \mathcal{L}_{\sla{P},\,\sla{T}}
    = 
        - \sum_{q=\text{all}} \text{Im}(m_q) \bar{q} i \gamma_5 q 
        + 
        \theta_G \frac{\alpha_s}{8\pi} G_{\mu\nu}^a\tilde{G}^{a\mu\nu} \,,
    \label{eq:thetaG}
\end{align}
where $m_q \equiv |m_q|\exp({i\theta_q})$ is the complex quark masses, $G^a_{\mu \nu}$ is the gluon field strength tensor, $\tilde{G}^{a \mu \nu} \equiv \frac{1}{2} \epsilon^{\mu \nu \rho \sigma} G^a_{\rho \sigma}$ is its dual (with $\epsilon^{0123} = +1$),
and $\alpha_s = g_s^2 / 4 \pi$ with the $SU(3)_C$ coupling constant $g_s$.
The anomalous axial Ward identity yields a rephasing-invariant combination of phases, known as the QCD $\theta$ angle $\bar{\theta} = \theta_G - \sum_q \theta_q$ \cite{Adler:1969gk, Adler:1969er, Bell:1969ts, Fujikawa:1979ay}.
The experimental upper bound on the electric dipole moment (EDM) of neutron \cite{Abel:2020pzs}, combined with lattice calculations \cite{Liang:2023jfj}, imposes a stringent constraint $|\bar{\theta}| \lesssim 10^{-10}$, nevertheless its natural value is expected to be $O(1)$ because there is no symmetry in the SM to suppress it.

Several new physics models have been proposed to address the strong $CP$ problem. 
One of the leading solutions is the Peccei-Quinn (PQ) mechanism, which dynamically selects a vacuum where  $\bar{\theta}=0$~\cite{Peccei:1977hh, Weinberg:1977ma, Wilczek:1977pj}. 
Other promising models involve extending $P$ \cite{Beg:1978mt, Mohapatra:1978fy, Babu:1989rb, Barr:1991qx} or $CP$ \cite{Nelson:1983zb, Barr:1984qx, Barr:1984fh} symmetries 
based on the $CPT$ theorem.
These symmetries forbid a tree-level $\theta$ term, but since they must be spontaneously broken at low energies (where the SM is valid), $\bar\theta$ can still be generated radiatively. Therefore, evaluating quantum corrections to 
$\bar\theta$ is essential.

In many such models, radiative corrections to the quark mass phases --linked to the chiral anomaly-- have traditionally been the only considered contributions to $\bar\theta$ \cite{Ellis:1978hq, Babu:1989rb, Hall:2018let, Craig:2020bnv, deVries:2021pzl}.  In contrast, some of the present authors previously proposed a method to directly compute the radiative corrections to $\bar \theta$ using a diagrammatic approach \cite{Hisano:2023izx,Banno:2023yrd} based on the Fock-Schwinger gauge \cite{Fock:1937dy, Schwinger:1951nm, key86364857, Cronstrom:1980hj, Shifman:1980ui, Dubovikov:1981bf, Novikov:1983gd}.
By applying this method to toy models with $CP$-violating Yukawa couplings  \cite{Banno:2023yrd}, they showed that two-loop corrections to $\bar \theta$ arise not only from quark mass phases but also from operator mixing via quark chromo-electric dipole moment operators generated at the one-loop level.  
 This is consistent with renormalization-group (RG) analyses in the Standard Model Effective Field Theory (SMEFT) \cite{Jenkins:2017dyc} and studies of CKM contributions to $\bar\theta$ \cite{Khriplovich:1985jr, Khriplovich:1993pf, Pospelov:1994uf}.

    In our earlier work \cite{Banno:2023yrd}, we found that ultraviolet (UV) divergences from two-loop diagrams are renormalized by fermion mass counterterms, yielding results consistent with $4$-dimensional EFT as mentioned above. 
    However, a shortcoming remained in the treatment of $\gamma_5$.
    We employed dimensional regularization (DR), but did not treat the $\gamma$-trace consistently.
    The relevant two-loop diagrams contain traces,
    such as $\opn{tr} [(\prod_{j=1}^6  \gamma_{\mu_j}) \gamma_5]$, while we evaluated the trace as in four dimensions to introduce the Levi-Civita symbol.
      
    The BMHV scheme \cite{tHooft:1972tcz, Bollini:1972ui, Breitenlohner:1977hr}, established by Breitenlohner, Maison, 't~Hooft and Veltman, provides a consistent definition of  $\gamma_5$ (and also the Levi-Civita symbol $\epsilon^{\mu\nu\rho\sigma}$).\footnote{If the BMHV scheme is adopted, 
    the gauge symmetry-breaking terms are introduced in chiral gauge theories (see, for example, Ref.~\cite{Belusca-Maito:2020ala}). On the other hand, since QCD is a vector-like theory, such   
    gauge symmetry-breaking terms are not needed in this paper.}.
    When the $d$-dimensional trace is treated properly by adopting the BMHV scheme, the equivalence between the two-loop results and the $4$-dimensional EFT might not be automatically guaranteed.
    In the BMHV scheme, full $d$-dimensional Lorentz invariance is broken since $\gamma_5$ only anticommutes with the $4$-dimensional $\gamma$ matrices. 
    This necessitates the introduction of evanescent operators—terms that vanish in the limit $d\rightarrow 4$ but are needed for consistency in dimensional regularization.
    Hence, modifications sourced by the BMHV scheme are not only the trace calculation but also the introduction of the evanescent operators.

\begin{comment}
In this paper, we investigate a one-flavor QCD toy model with a real scalar field and a $CP$-violating Yukawa interaction that breaks $P$ and $T$ symmetries. 
We compute the two-loop radiative corrections to $\bar\theta$ using the BMHV scheme and the Fock-Schwinger gauge. 
A new dimensionless coupling $\xi$ is introduced for an evanescent kinetic term for a fermion that breaks chiral symmetry:
\begin{eqsp}
    {\cal L}
    \supset 
        (1 + \xi)
        \bar{q}_L i \hat{\sla{D}} q_R
        +
        {\rm h.c.} \,.
\end{eqsp}
where $\xi$ is complex. We find that this evanescent kinetic term contributes to $\bar\theta$ at one-loop level, making the effective bare angle $\bar{\theta}_G\equiv \theta_G+{\rm Im}[\xi]$. At two-loop level, $\theta_G$
becomes renormalization-scale dependent in the toy model, while $\bar{\theta}_G$
remains renormalization-scale independent. Moreover, the unphysical $CP$-odd rephasing-invariants appear in the radiative corrections to $\bar \theta$ at two-loop level due to the evanescent kinetic term, though they can be consistently removed by appropriately choosing $\xi$. These results match those from Ref.~\cite{Banno:2023yrd}  and support the BMHV scheme as a valid method for computing higher-order radiative corrections to $\bar\theta$.
\end{comment}

    In this paper, we investigate a one-flavor QCD toy model with a real scalar field and a $CP$-violating Yukawa interaction that breaks $P$ and $T$ symmetries. 
    We compute the two-loop radiative corrections to $\bar\theta$ using the BMHV scheme and the Fock-Schwinger gauge. 
    The trace of the $\gamma$ matrices with the BMHV scheme holds not only the $4$-dimensional part but also an evanescent part, which is not included in the previous work \cite{Banno:2023yrd}.
    On the other hand, a new dimensionless coupling $\xi$ is introduced for an evanescent kinetic term for the fermion that breaks chiral symmetry:
    \begin{eqsp}
        {\cal L}
        \supset 
            (1 + \xi)
            \bar{q}_L i \hat{\sla{D}} q_R
            +
            {\rm h.c.} \,.
    \end{eqsp}
    where $\xi$ is complex. 
    We find that this evanescent kinetic term contributes to $\bar\theta$ at the one-loop level, and both the $(d-4)$-dimensional part of the $\gamma$-trace and the evanescent kinetic-term contribution do not cancel each other.
    The remaining terms can be removed by renormalization of $\theta_G$ properly, %so that
    though $\theta_G$ becomes renormalization-scale dependent in the toy model.
    We show its scale dependence at the two-loop level is canceled by that of $\xi$ after renormalization of $\theta_G$.
    Then, the obtained QCD $\theta$ angle is consistent with the previous work \cite{Banno:2023yrd}.
%    A part of the scale-dependence is canceled by one of $\xi$, but the remaining part will be done by one more higher order perturbative corrections, which is out of our current scope.

Finally, we examine the rephasing invariance of $\bar\theta$ in this framework. In the Fujikawa method, anomalous Jacobians for fermions appear under chiral rotations. However, in dimensional regularization, the role of such Jacobians is less clear. Our results suggest that the renormalized $\theta_G$ is correctly transformed under the chiral rotation due to the property of the evanescent kinetic term and 
that the Jacobians for fermions are trivial in the BMHV scheme.

The rest of the paper is organized as follows.
In Sec.~\ref{sec:two-loop}, we show the one-flavor QCD Lagrangian, incorporating a $CP$-violating Yukawa interaction.  
The evanescent kinetic term with the coefficient $\xi$ is introduced in the BMHV scheme there. 
We introduce the rephasing-invariants in the original Lagrangian and also the unphysical rephasing-invariants in the BMHV scheme. In Sec.~\ref{sec:re-parameterization}, we show the result for the diagrammatic calculation of the QCD $\theta$ angle at the two-loop level with the Fock-Schwinger gauge.
In Sec.~\ref{sec:re-parameterization}, we discuss the renormalization of $\theta_G$ and $\xi$, and show that our result for the radiative correction to the QCD $\theta$ angle is consistent with our previous results derived in Ref.~\cite{Banno:2023yrd}. 
Section~\ref{sec:conclusion} is devoted to the conclusion and discussion. There, we discuss the rephasing invariance of the radiative corrections to the QCD $\theta$ angle.

%=======================================================
%        Sec.2 $d$-dimensional Lagrangian and evanescent contribution to the QCD $\theta$ angle
%=======================================================
\section{Lagrangian in BMHV scheme}
\label{sec:two-loop}

Let us begin our discussion with 
the one-flavor QCD Lagrangian with a real scalar, where the $P$ and $T$ symmetries are not imposed, and a $CP$-violating Yukawa interaction is introduced. This toy Lagrangian is
\begin{align}
\begin{aligned}
    {\cal L}
    &=
        \frac12 (\partial_\mu \phi)^2
        -
        \frac{1}{2} m_\phi^2 \phi^2
        -
        \frac{1}{4} {G}^a_{\mu \nu} {G}^{a\mu \nu}
        +
        \theta_G \frac{\alpha_s}{8 \pi}
        G^a_{\mu \nu} \tilde{G}^{a \mu \nu}
      \\
       &\quad  +
        \bar{q} i \sla{D} q
        -
        \bar{q} 
        \left[
            \opn{Re} (m_q)
            +
            \opn{Im} (m_q) i \gamma_5
        \right]
        q
        -
        (y_q \bar{q} P_R q \phi
        +
        {\rm h.c.}) \,,
    \label{eq:4 Lag}
\end{aligned}    
\end{align}
where $D_\mu = \partial_\mu + i g_s T_s^a G_\mu^a$ and $P_{R/L}=(1\pm\gamma_5)/2$. Here, $g_s$, $G_\mu^a$, ${G}^a_{\mu \nu}$ and $T_s^a$ denote %are for
the $SU(3)_c$ gauge coupling constant, gauge bosons,  field-strengths, and generators, respectively.
For simplicity, we restrict the mass parameter as $\opn{Im} (m_q) \ll \opn{Re} (m_q)$ so that the $\gamma_5$-mass term can be treated perturbatively.
We also assume that $m_q$ is larger than $\Lambda_{\rm QCD}$ so that the perturbation works. 

Observables should be independent of the rephasing of fields in the Lagrangian.
It means that the QCD $\theta$ angle has to be composed of $CP$-odd  rephasing-invariants of parameters in Lagrangian, 
even after including the radiative corrections. For simplicity, we explain the rephasing-invariants in the Lagrangian in Eq.~\eqref{eq:4 Lag}, assuming the Fujikawa method.
The chiral rotation generates the anomalous Jacobian due to the chiral anomaly, and it  gives a correction to the QCD $\theta$ angle. Thus, $\theta_G - \opn{arg} [m_q]$ is rephasing-invariant \cite{Adler:1969gk, Adler:1969er, Bell:1969ts, Fujikawa:1979ay}.  A product $m_qy^\star_q$ is also rephasing-invariant. However, 
under a $Z_2$ symmetry under which $y_q\rightarrow -y_q$ and $\phi\rightarrow -\phi$,  the corrections to the QCD $\theta$ angle are proportional to the even power of the Yukawa coupling \cite{Banno:2023yrd}. Thus, the correction to the QCD $\theta$ angle should be proportional to $\im{(y_q m_q^*)^2}$. It becomes $\im{y_q^2} \re{m_q^2}$ under the assumption $\re{m_q} \gg \im{m_q}$. The two-loop corrections 
to the QCD $\theta$ angle, evaluated in Ref.~\cite{Banno:2023yrd}, are proportional to it.

\begin{figure}
    \centering
    \includegraphics[width=0.9\linewidth]{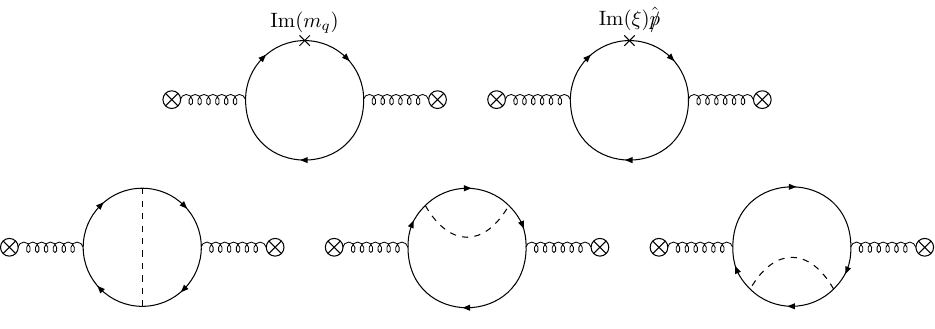}
    \caption{
        Loop diagrams that contribute to the QCD $\theta$ angle.
        The top two diagrams give the one-loop contribution, $\delta \theta^{(1)}$.
        The bottom three diagrams produce the two-loop contributions, where we name each diagram as $\delta \theta^{(2)}_a, \delta \theta^{(2)}_b$, and $\delta \theta^{(2)}_c$ from left to right.
    }
    \label{fig:loop diagrams}
\end{figure}

Now we extend the Lagrangian to $d$ dimensions in order to evaluate the loop diagrams represented in Fig.~\ref{fig:loop diagrams} with dimensional regularization.
Here, we adopt the BMHV scheme for the $\gamma_5$ treatment in the $d$-dimensional loop integrals \cite{tHooft:1972tcz,Breitenlohner:1977hr,Bollini:1972ui}.
Hence, $\gamma_5$ and $\epsilon^{\mu \nu \rho \sigma}$ are not modified with the dimensional regularization, and the algebra of the $\gamma$ matrix is defined as
\begin{align}
    \gamma_5 
    =
        - \frac{i}{4 !}
        \bar{\epsilon}^{\mu \nu \rho \sigma}
        \gamma_\mu \gamma_\nu \gamma_\rho \gamma_\sigma \,,&
    \qquad
    \hat{\epsilon}^{\mu \nu \rho \sigma} 
    = 
        0\,, 
\end{align}
and
\begin{align}
    \left\{ 
        \bar{\gamma}_\mu, \gamma_5
    \right\} 
    =
        0\,,&
    \qquad
    \left[
        \hat{\gamma}_\mu, \gamma_5
    \right] 
    =
        0 \,.
\end{align}
Here, the bar and hat on a Lorentz tensor/vector mean that they are $4$- and $(d-4)$-dimensional quantities, respectively.
The Lagrangian in the BMHV scheme is expressed as
\begin{align}
\begin{aligned}
   {\cal L}&=
        \frac12 (\bar{\partial}_\mu \phi)^2
        +
        \frac12 (\hat{\partial}_\mu \phi)^2
        -
        \frac{1}{2} m_\phi^2 \phi^2
        -
        \frac{1}{4} G^a_{\mu \nu} G^{a\mu \nu}
        +
        \theta_G \frac{\alpha_s}{8 \pi}
        \bar{G}^a_{\mu \nu} \tilde{\bar{G}}^{a \mu \nu}
        \\ 
    &\quad +
        \bar{q} i \bar{\sla{D}} q
        +
        \left[(1 + \xi)
        \bar{q}_L i \hat{\sla{D}} q_R+{\rm h.c.}\right]
        -
        \bar{q} 
        \left[
            \opn{Re} (m_q)
            +
            \opn{Im} (m_q) i \gamma_5
        \right]
        q\\
&        -
        \left(y_q \bar{q} P_R q \phi
        +
        {\rm h.c.}\right)\,.
    \label{eq:d Lag}
\end{aligned}    
\end{align}
We introduce the evanescent kinetic term of fermion with the coupling constant $1 + \xi$, where $\xi$ is a complex number.\footnote{The evanescent kinetic term was introduced in Ref.~\cite{Schubert:1988ke}, which RG equations in the Yukawa model were derived at the two-loop level. } 
Such a kinetic term does not appear when  we assume the anticommutation relation $\{ \gamma_\mu, \gamma_5 \} = 0$ holds in arbitrary $d$ dimensions, as in Ref.~\cite{Banno:2023yrd}.
%        , and hence, the evanescent kinetic term and its contributions inevitably fail to be included.
The evanescent coupling constant is not necessarily the same as the $4$-dimensional kinetic term, but the unity is needed for producing the ordinary fermion propagator. 
Hence, we require an assumption $|\xi| \ll 1$.
It is not proper parameterization to set $\xi = 0$, because it varies with the renormalization scale differently from the $4$-dimensional kinetic term, which is discussed in the next section.
The evanescent kinetic term breaks the chiral symmetry so that it is similar to the mass term of fermion. On the other hand.
the kinetic terms for the scalar and gauge bosons are taken to be $d$-dimensional Lorentz invariant here. This is because 
the violation of the $d$-dimensional Lorentz invariance
in the kinetic terms contributes to the QCD $\theta$ angle in more than the two-loop level assuming that the evanescent kinetic terms are induced by the radiative corrections.

Before calculating the correction to the QCD $\theta$ angle, we discuss the  $CP$-odd rephasing-invariants in the Lagrangian in the BMHV scheme given in Eq.~\eqref{eq:d Lag}. 
New rephasing invariants appear due to the introduction of the evanescent kinetic term, 
\begin{equation}
    m_q (1 + \xi)^\star\, ,
    \quad
    y_q (1 + \xi)^\star \,.
\end{equation}
Here, the coefficient of the evanescent kinetic term include 1 so that the rephasing transformation of the Lagrangian parameters is obvious. 
We show in the next section that the correction to the QCD $\theta$ angle, 
proportional to $\im{\xi}$, appears at the one-loop level. 
It comes from the rephasing-invariant 
$m_q (1 + \xi)^\star$ ($\simeq -\re{m_q}\im{\xi}$ for 
$\im{m_q}\ll\re{m_q}$). At the two-loop level, 
the correction proportional to $\im{(y_q (1 + \xi)^\star)^2}$ $(\simeq \im{y_q^2}$ for $|\xi|\ll1$), which is $Z_2$ even under $y_q\rightarrow -y_q$ and $\phi\rightarrow -\phi$,  appears. Since they are unphysical, they should be consistently removed.

\section{Radiative corrections to QCD $\theta$ angle in BMHV scheme}

Now, we evaluate the radiative corrections to the QCD $\theta$ angle in the BMHV scheme up to the two-loop level by the diagrammatic calculation with the Fock-Schwinger gauge method \cite{Hisano:2023izx,Banno:2023yrd} .
The calculation details for each diagram are shown in Ref.~\cite{Hisano:2023izx, Banno:2023yrd}; therefore, we do not present them here and instead show only the results.
The one-loop contribution $\delta \theta^{(1)}$ and the two-loop contributions $\delta \theta^{(2)}_a, \delta \theta^{(2)}_b$, and  $\delta \theta^{(2)}_c$ in Fig.~\ref{fig:loop diagrams} are 
\begin{align}
    & \begin{aligned}
        \delta \theta^{(1)}
        =
            - \frac{\opn{Im} [m_q]}{\opn{Re} [m_q]}
            -
            \frac{2 (d-4)}{d}
            \im{\xi}
            \left[ 
                \frac{1}{\epsilon}
                -
                \frac{1}{2}
                +
                \log \frac{Q^2}{(\re{m_q})^2}
            \right] \,,
    \end{aligned}
    \\
    & \begin{aligned}
        \delta \theta^{(2)}_a
        &=
            \left(
                \frac{\alpha_s}{8 \pi} (G \tilde{G})
            \right)^{-1}
            \frac{g_s^2}{2} (G \tilde{G})
            \int_{p,q}
            \frac{
                \opn{Im} [y_q^2]
                \left[
                    (\opn{Re} [m_q])^2 + \hat{p} \cdot \hat{q}
                \right]
            }{
                [p^2 - (\opn{Re} [m_q])^2]^2
                [q^2 - (\opn{Re} [m_q])^2]^2
                [(p-q)^2 - m_\phi^2]
            }
            \\
        &=
            \frac{\opn{Im} [y_q^2]}{16 \pi^2}
                (\opn{Re} [m_q])^2
                I^{(2,2,1)}( (\Re[m_q])^2, m_\phi^2 )
            \\
            &- \frac{d-4}{2d}
            \frac{\opn{Im} [y_q^2]}{16 \pi^2}
            \left[   
                    I^{(2,2,0)}
                    -
                    I^{(1,2,1)}
                    -
                    I^{(2,1,1)}
                    +
                    \left( m_\phi^2 - 2 (\opn{Re} [m_q])^2 \right)
                    I^{(2,2,1)}
            \right] ( (\Re[m_q])^2, m_\phi^2 )\,,
        \label{eq:delta theta 2a}
    \end{aligned}
    \\
    & \begin{aligned}
        &
            \delta \theta^{(2)}_b + \delta \theta^{(2)}_c
            \\
        &=
            \left(
                \frac{\alpha_s}{8 \pi} (G \tilde{G})
            \right)^{-1}
            g_s^2 (G \tilde{G})
            \int_{p,q}
            \frac{
                \opn{Im} [y_q^2]
                \left[
                    (\opn{Re} [m_q])^2 + \hat{p} \cdot \hat{q}
                \right]
            }{
                [p^2 - (\opn{Re} [m_q])^2]^3
                [q^2 - (\opn{Re} [m_q])^2]
                [(p-q)^2 - m_\phi^2]
            }
            \\
        &=
            \frac{\opn{Im} [y_q^2]}{16 \pi^2}   
                (\opn{Re} [m_q])^2
                2 I^{(3,1,1)} ( (\Re[m_q])^2, m_\phi^2 )
            \\
           & -
\frac{2(d-4)}{2d} \frac{\opn{Im} [y_q^2]}{16 \pi^2}                      
                \left[
                    I^{(3,1,0)}
                    -
                    I^{(2,1,1)}
                    -
                    I^{(3,0,1)}
                    +
                    \left( m_\phi^2  - 2 (\opn{Re} [m_q])^2 \right)
                    I^{(3,1,1)}
                \right]
             ( (\Re[m_q])^2, m_\phi^2 ) \,,
        \label{eq:delta theta 2bc}
    \end{aligned}
\end{align}
where the two-loop functions $I^{(l,m,n)} (x,z)$ are defined as
\begin{equation}
    I^{(l,m,n)} (x, z)
    \equiv
        \int
        \frac{d^d p \, d^d q}{(2 \pi)^{2d}}
        \frac{1}{
            [p^2 - x]^l
            [q^2 - x]^m
            [(p-q)^2 - z]^n
        } \,.
\end{equation}
We also use $Q^2\equiv 4\pi\mu^2{\rm e}^{-\gamma_E} $ ($\mu$ is the renormalization scale in dimensional regularization. 
Note that all renormalization scale dependence can be represented by $Q^2$ in $\overline{\rm MS}$ scheme.). 

It is found that there are two additional contributions in the  BMHV scheme compared with the result of Ref.~\cite{Banno:2023yrd}. They are induced by the evanescent kinetic term, and they are proportional to the unphysical rephasing invariants discussed in the previous section. 
First, the contribution proportional to ${\rm Im}[\xi]$ is generated at the one-loop level ($\delta \theta^{(1)}$). Second is
the second terms in the second lines of Eqs.~\eqref{eq:delta theta 2a} and \eqref{eq:delta theta 2bc}, which are  proportional to ${\rm Im}[y_q^2]$.  Inner products of the $(d-4)$ dimensional loop momenta $\hat{p}\cdot\hat{q}$, appear at the numerators of the integrands due to the evanescent kinetic term.

The new contributions in the  BMHV scheme are superficially proportional to $(d-4)$, though they
give not only finite but also infinite contributions in the limit of $d\to 4$. This is because  
\begin{align} 
    - \frac{2 (d-4)}{d}
    \im{\xi}
    \left[ 
        \frac{1}{\epsilon}
        -
        \frac{1}{2}
        +
        \log \frac{Q^2}{(\re{m_q})^2}
    \right]
    &=
        \im{\xi} 
        +
        \mathcal{O} (\epsilon^1) \,,
        \label{eq:evanescent div0}
        \\
    I^{(2,2,0)}
    -
    I^{(1,2,1)}
    -
    I^{(2,1,1)}
    +
    (m_\phi^2 - 2 m_q^2)
    I^{(2,2,1)}
    &=
        \frac{1}{\epsilon}
        +
        \mathcal{O} (\epsilon^0) \,,
        \label{eq:evanescent div1}
        \\
    I^{(3,1,0)}
    -
    I^{(2,1,1)}
    -
    I^{(3,0,1)}
    +
    (m_\phi^2 - 2 m_q^2)
    I^{(3,1,1)}
    &=
        \frac{1}{2 \epsilon^2}
        +
        \frac{1}{\epsilon}
        \log \frac{Q^2}{(\re{m_q})^2}
        +
        \mathcal{O} (\epsilon^0) \,.
        \label{eq:evanescent div2}
\end{align}
Equation \eqref{eq:evanescent div0} appears in $\delta\theta^{(1)}$, and Eqs.~\eqref{eq:evanescent div1} and \eqref{eq:evanescent div2} are included in $\delta\theta_a^{(2)}$ 
and $\delta\theta_b^{(2)}+\delta\theta_c^{(2)}$, respectively.
Note that the last one  
gives the UV-divergent  contribution to $\delta\theta_b^{(2)}+\delta\theta_c^{(2)}$ (\eq{eq:delta theta 2bc}).

%=======================================================
%        Sec.3 Evanescent free QCD theta angle
%=======================================================
\section{Removal of unphysical contributions to QCD \texorpdfstring{\boldmath{$\theta$}}{theta} angle}
\label{sec:re-parameterization}

The radiative corrections to the QCD $\theta$ angle, obtained in the previous section,  have UV divergences, so we need to renormalize them.  
In \eq{eq:delta theta 2bc}, both the first and second terms are divergent. The second term is the contribution from the evanescent kinetic term, as mentioned in the previous section. In Ref.~\cite{Banno:2023yrd}, it is argued that the divergence in the first term comes from the fermion two-point function, and it can be subtracted by the mass counterterm. 
In this paper, we adopt the BMHV scheme. When we calculate the fermion two-point diagrams in the BMHV scheme, they should contribute to not only the mass term but also the evanescent kinetic term. Hence, we decompose the fermion two-point function into three parts: the mass term $i \Sigma_m$, the $4$-dimensional kinetic term $i \Sigma_{\bar{2}}$ and the evanescent one $i \Sigma_{\hat{2}}$. The fermion-scalar one-loop diagram contributes to them as
\begin{align}
     i \Sigma_m P_R
    &=
        \frac{i}{16 \pi^2}
        y_q^2 \opn{Re} [m_q] P_R
        \left[ 
            \frac{1}{\epsilon}
            -
            F_0 (p^2, (\opn{Re} [m_q])^2, m_\phi^2)
        \right] \,,
        \\
     i \Sigma_{\bar{2}}
    &=
        \frac{i}{16 \pi^2}
        | y_q |^2 \bar{\sla{p}}
        \left[ 
            \frac{1}{2 \epsilon}
            -
            F_1 (p^2, (\opn{Re} [m_q])^2, m_\phi^2)
        \right] \,,
        \\
     i \Sigma_{\hat{2}} P_R
    &=
        \frac{i}{16 \pi^2}
        y_q^2 \hat{\sla{p}} P_R
        \left[ 
            \frac{1}{2 \epsilon}
            -
            F_1 (p^2, (\opn{Re} [m_q])^2, m_\phi^2)
        \right] \,,
\end{align}
where
\begin{align}
    F_0 (p^2, x_1, x_2)
    &=
        \int_0^1 d z
        \log 
        \frac{
            - z (1-z) p^2
            +
            (1-z) x_1
            + 
            z x_2 
        }{Q^2} \,,
        \\
    F_1 (p^2, x_1, x_2) 
    &=
        \int_0^1 d z \,
        z
        \log 
        \frac{
            - z (1-z) p^2
            +
            (1-z) x_1
            + 
            z x_2 
        }{Q^2} \,.
\end{align}
The radiative correction to $i \Sigma_{\bar{2}}$  is  $CP$ even, while that to $i \Sigma_{\hat{2}}$, may depend on the $CP$-violating phase of the Yukawa coupling due to the chiral symmetry breaking.
Thus, we cannot subtract those divergences by the ordinary perturbative renormalization because the ordinary wave-function renormalization introduces the wave-function renormalization factor $\bar{q}^0 i \sla{\partial} q^0 = Z_2 \bar{q} i \sla{\partial} q$ where the subscript $0$ indicates a bare quantity.
However, we have already introduced the coupling constant for the evanescent kinetic term $\xi$.
Therefore, an additional renormalization factor for the evanescent kinetic term can be prepared.
The counterterms for the mass term and both of the kinetic terms are defined as
\begin{align}
    \mathcal{L}
    &\supset
        \bar{q}^0 i \bar{\sla{\partial}} q^0
        +
        \left[
            (1+\xi^0) \bar{q}^0_L i \hat{\sla{\partial}} q^0_R
            +
            {\rm h.c.}
        \right]
        -
        (
            m_q^0 \bar{q}_L^{0} q_R^{0}
            +
            {\rm h.c.}
        )
      \nonumber  \\
    &=
        \bar{q}i \sla{\partial} q
        -
        (
            m_q \bar{q}_L q_R
            +
            {\rm h.c.}
        )
     \\
    &\quad 
        +
        \delta_{\bar{2}} 
            \bar{q} i \bar{\sla{\partial}} q
        +
        \left[
            \left( \xi (Q) + \delta_{\hat{2}} \right)
            \bar{q}_L i \hat{\sla{\partial}} q_R
            +
            {\rm h.c.}
        \right]
        -
        (
            \delta_{m} m_q \bar{q}_L q_R
            +
            {\rm h.c.}
        ) \,.\nonumber 
\end{align}
We then determine these counterterms using the $\overline{\rm MS}$ scheme as follows:
\begin{align}
     \delta_m
    &=
        \frac{1}{16 \pi^2} y_q^2
        \frac{1}{\epsilon}\,,
        \label{eq:counterterm m}
        \\
     \delta_{\bar{2}}
    &=
        - \frac{1}{16 \pi^2} \frac{1}{2} |y_q|^2
        \frac{1}{\epsilon} \,,
        \label{eq:counterterm bar2}
        \\
     \delta_{\hat{2}}
    &=
        - \frac{1}{16 \pi^2} \frac{1}{2} y_q^2
        \frac{1}{\epsilon} \,.
        \label{eq:counterterm hat2}
\end{align}

As a result, the effective evanescent kinetic term becomes 
\begin{equation}
    \mathcal{L}^\text{eff}
    \supset
        \left( 
            1 
            +
            \xi (Q) 
            -
            \frac{y_q^2}{16 \pi^2}
            F_1 (p^2, m_q^2, m_\phi^2)
        \right)
        \bar{q}_L i \hat{\sla{\partial}} q_R
        +
        {\rm h.c.} .
        \label{eq:Lkinhat}
\end{equation}
%and then, assuming that Eq.~\eqref{eq:Lkinhat} is $Q$ independent, $\xi (Q)$ can be obtained as $\xi (Q) =\xi (Q_0) -\frac{y_q^2}{32 \pi^2}\log \frac{Q^2}{Q_0^2} $ by solving the RG equation 
Equation~\eqref{eq:Lkinhat} is $Q$ independent by definition of the $\overline{\rm MS}$ scheme, and then $\xi (Q)$ can be obtained as $\xi (Q) =\xi (Q_0) -\frac{y_q^2}{32 \pi^2}\log \frac{Q^2}{Q_0^2} $ by solving the RG equation
\begin{equation}
Q\frac{d}{dQ}\xi(Q)=-\frac{y_q^2}{16\pi^2}    \,,
    \label{eq:xi RG solution}
\end{equation}
with a boundary condition at $Q_0$.

The counterterm corrections to the QCD $\theta$ angle can be obtained by the top diagrams in Fig.~\ref{fig:loop diagrams} by replacing $\opn{Im} [m_q]$ and $\opn{Im} [\xi]$ into its counterterms, respectively,
\begin{eqsp}
    \delta \theta_{\rm c.t.}^{(2)}
    &=
        -
        \opn{Im} [\delta_{m}]
        -
        \frac{2 (d-4)}{d}
        \opn{Im} [\delta_{\hat{2}}]
        \left\{
            \frac{1}{\epsilon}
            -
            \frac{1}{2}
            +
            \log \frac{Q^2}{(\opn{Re} [m_q])^2}
        \right\} \,.
\end{eqsp}
Furthermore, we consider the renormalization of $\theta_G$, 
since the above counterterms cannot absorb all divergences in the two-loop diagrams, as will be shown below. The bare parameter for $\theta_G$ is given as 
\begin{eqsp}
    \theta^{0}_G 
    =
        \mu^{-2 \epsilon} 
        \left[\theta_G (Q) + \delta_{\theta_G}\right]\, .
        \label{eq:theta_G bare}
\end{eqsp}

Below, we combine all contributions to the QCD $\theta$ angle as
\begin{align}
    &
        \theta_G (Q)
        +
        \delta \theta^{(1)}
        +
        \delta \theta^{(2)}_a
        +
        \delta \theta^{(2)}_b
        +
        \delta \theta^{(2)}_c
        +
        \delta \theta^{(2)}_{\rm c.t.}
        +
        \delta_{\theta_G}
        \nonumber \\
    &\quad =
        \theta_G (Q)
        - 
        \frac{\opn{Im} [m_q]}{\opn{Re} [m_q]}
        +
        \left[ 
            \frac{\opn{Im} [y_q^2]}{16 \pi^2}
            (\opn{Re} [m_q])^2
            \left(
                I^{(2,2,1)}
                +
                2 I^{(3,1,1)}
            \right)
            -
            \opn{Im} [\delta_m]
        \right]
        \nonumber \\
    &\qquad 
        -
        \frac{d-4}{2d}
        \left[
            \frac{\opn{Im} [y_q^2]}{16 \pi^2}
            \left\{
                I^{(2,2,0)}
                -
                I^{(1,2,1)}
                -
                3 I^{(2,1,1)}
                +
                2 I^{(3,1,0)}
                -
                2 I^{(3,0,1)}                
            \right.
        \right.
       \nonumber \\
    & \quad \qquad 
            +
            \left.
                \left( m_\phi^2 - 2 (\opn{Re} [m_q])^2 \right)
                \left(
                    I^{(2,2,1)} 
                    +
                    2 I^{(3,1,1)}
                \right)
            \right\}
      \nonumber  \\
    &\quad \qquad +
        \left.
            4
            \opn{Im} [\xi (Q) + \delta_{\hat{2}}]
            \left\{
                \frac{1}{\epsilon}
                -
                \frac{1}{2}
                +
                \log \frac{Q^2}{(\opn{Re} [m_q])^2}
            \right\}
        \right]
        +
        \delta_{\theta_G}
        \nonumber\\ 
    &\quad =
        - 
        \frac{\opn{Im} [m_q]}{\opn{Re} [m_q]}
        +
        \frac{\opn{Im} [y_q^2]}{16 \pi^2}
        (\opn{Re} [m_q])^2
        \left(
            I^{(2,2,1)}
            +
            2 \bar{I}^{(3,1,1)}
        \right)
       \nonumber \\ 
    &\qquad
        +
        \left[
            \theta_G (Q)
            +
            \im{\xi (Q)}
            +
            \frac{\im{y_q^2}}{16 \pi^2}
            \frac{5}{8}
        \right]
        +
        \left[
            -
            \frac{\im{y_q^2}}{16 \pi^2}
            \frac{1}{4 \epsilon}
            +
            \delta_{\theta_G} 
        \right]
        +
        O(\epsilon^1)\,.
        \label{eq:total_theta}
\end{align}
The first and second terms in the last line are the contributions equivalent to the result of Ref.~\cite{Banno:2023yrd}, and $\bar{I}^{(3,1,1)}$ is the renormalized two-loop function defined there.
%The $\epsilon$ pole remains in the last line of q{eq:total_theta}, and it is subtracted by the counterterm $\delta_{\theta_G}$ determined as
The evanescent contributions which are proportional to $\im{y_q^2}$ can be seen in the third and forth terms of the last line of Eq.~\eqref{eq:total_theta}, so it means that the renormalization of the evanescent kinetic term does not cancel completely the $(d-4)$-dimensional contributions sourced from $\delta \theta_a^{(2)} + \delta \theta_b^{(2)} + \delta \theta_c^{(2)}$.
We determine the counterterm $\delta_{\theta_G}$ to subtract the evanescent divergence as
\begin{eqsp}
    \delta_{\theta_G}
    =
        \frac{\im{y_q^2}}{16 \pi^2}
        \frac{1}{4 \epsilon} \,.
\end{eqsp}
This implies that the $\theta_G(Q)$ depends on the renormalization scale $Q$. The RG equation for $\theta_G$ is
\begin{eqsp}
    Q \frac{d}{d Q} \theta_G (Q)
    &=
        2 \epsilon (\theta_G + \delta_{\theta_G})
        - 
        Q \frac{d}{d Q}
        \delta_{\theta_G}
        =
        \frac{\im{y_q^2}}{16 \pi^2} \,.
        \label{eq:theta RG equation}
\end{eqsp}
This equation is correct up to the two-loop level.
Here, we used the fact that the bare parameter for the Yukawa coupling $y_q$ is given by 
\begin{eqsp}
    y^{0}_q 
    =
        \mu^{\epsilon} 
        \left[y_q (Q) + \delta_{y_q}\right]\, .
\end{eqsp}

We define an effective $\theta$ angle, 
\begin{equation} 
    \theta'^{(\textrm{4-dim})}_G
    \equiv
        \theta_G (Q) 
        + 
        \im{\xi (Q)} 
        +
        \frac{\im{y_q^2}}{16 \pi^2}
            \frac{5}{8}\,.
    \label{eq:effective theta}
\end{equation}
From Eqs.~\eqref{eq:xi RG solution} and \eqref{eq:theta RG equation}, we find that $ \theta_G+\im{\xi (Q)}$ is RG-invariant at the two-loop level, and then the effective $\theta$ angle, $ \theta'^{(\textrm{4-dim})}_G$, is also RG-invariant at the two-loop level. 
The last term in $\theta'^{(\textrm{4-dim})}_G$, proportional to ${\im{y_q^2}}/{(16 \pi^2)}$, might induce the scale dependence, though it is of higher order. We expect 
that the scale dependence vanishes when the higher order corrections are calculated.
%\ocom{
%    On the other hand, we find that the first term of \eqref{eq:theta RG equation} is the same as $\xi$'s RG equation from Eq.~\eqref{eq:xi RG solution},
%}
%\begin{eqsp}
%    Q \frac{d}{d Q} 
%    \theta'^{(\textrm{4-dim})}_G
%    =
%        0 \,.
%        \label{eq:RGEresult}
%        \ocom{modified}
%\end{eqsp}
%    This means the combination is scale independent up to two-loop order.
%    Its scale-dependence might appears from the higher order, but we expect that the scale-dependence can be removed by extra contributions which are given as two-loop diagrams Fig.~\ref{fig:loop diagrams} with insertion of $\xi$ in the fermion loop, which are effectively $(2+1)$-loop level.
%    This is because the CP-violating quantity $\im{y_q^2 (1 + \xi^*)^2} \simeq \im{y_q^2} + 2 \im{y_q^2 \xi^*}$ cannot be fully included in the above calculation, and the second term is generated by those $(2+1)$-loop diagrams.
%    Indeed, both Eq.~\eqref{eq:RGEresult} and those two-loop diagrams are proportional to $y_q^4$.
%    However, we do not explicitly calculate the two-loop diagrams here, since it is beyond the scope of this paper.
\begin{comment}
\rtext{[
Thus, we define the effective bare angle $\bar{\theta}_G$ as 
\begin{eqnarray}
    \bar{\theta}_G&=&\theta_G  + \im{\xi}. 
\end{eqnarray}
(remove)]
}
\end{comment}
As a result, the QCD $\theta$ angle up to the two-loop level is obtained as
\begin{eqsp}
    &
        \theta_G (Q)
        +
        \delta \theta^{(1)}
        +
        \delta \theta^{(2)}_a
        +
        \delta \theta^{(2)}_b
        +
        \delta \theta^{(2)}_c
        +
        \delta \theta^{(2)}_{\rm c.t.}
        +
        \delta_{\theta_G}
        \\
    &=
        - 
        \frac{\opn{Im} [m_q]}{\opn{Re} [m_q]}
        +
        \frac{\opn{Im} [y_q^2]}{16 \pi^2}
        (\opn{Re} [m_q])^2
        \left(
            I^{(2,2,1)}
            +
            2 \bar{I}^{(3,1,1)}
        \right)
        +
        \theta'^{(\textrm{4-dim})}_G \,.
\label{eq:final_eq}
\end{eqsp}
%\rtext{[
%    (remove)
%In the above equation, we still have the unphysical rephasing-invariant $\im{y_q^2}$. For example, the term can be removed by redefining $\bar{\theta}_G$. In the case, $\bar{\theta}_G$ becomes dependent on the renormaization scale though it is at higher order beyond our current calculation. 
%]}
    Thus, we get a result consistent with Ref.~\cite{Banno:2023yrd} under a redefinition of $\theta'^{(\textrm{4-dim})}_G$ as the purely $4$-dimensional bare angle of the $\theta$ term.

\section{Conclusions and discussion} 
\label{sec:conclusion}

In this paper, we evaluate the radiative corrections to the QCD $\theta$ angle in the simple toy model, in which the Yukawa coupling $y_q$ is $CP$-violating. 
We adopt the BMHV scheme in dimensional regularization in order to introduce $\gamma_5$ in a mathematically consistent manner.  
In the scheme, the chiral-symmetry breaking evanescent kinetic term for fermion with the coefficient $(1+\xi)$ is introduced. 
Its imaginary component $\im{\xi}$ gives the finite contributions to the QCD $\theta$ angle at the one-loop level in the limit of $d \to 4$ dimensions. 
In addition, the contribution proportional to $\im{y_q^2}$ to the QCD $\theta$ angle appears at the two-loop level from $(d-4)$-dimensional part of $\gamma$ traces in the limit of $d \to 4$ dimensions.
Those terms also come from the evanescent kinetic term. 
It is shown that the $\epsilon$ pole in the unphysical contributions is subtracted by renormalization of the bare $\theta$ angle, and its finite part is absorbed by definition of the effective $\theta$ angle, $\theta^{\prime (\textrm{4-dim})}_G$ (Eq.~\eqref{eq:effective theta}). 
We confirm that the angle does not depend on the renormalization scale up to the two-loop level.
We thus obtain a result for the radiative corrections to the QCD $\theta$ angle, which is consistent with Ref.~\cite{Banno:2023yrd}.

Last, we discuss the rephasing invariance in the QCD $\theta$ angle in the BMHV scheme. 
In $4$-dimensional space, the physical parameter of the QCD $\theta$ angle is defined as 
\begin{equation}
    \bar{\theta} \equiv \theta_G^{(\textrm{4-dim})} - \arg [m_q]\,,
    \label{eq:chiral anomaly}
\end{equation}
due to the chiral anomaly.
Here, $\theta_G^{(\textrm{4-dim})}$ is a coupling constant of the purely $4$-dimensional $\theta$ term.
$\bar{\theta}$ is a rephasing-invariant quantity.
Then %in $d$-dimension, 
we discuss the rephasing invariance of the obtained result, Eq.~\eqref{eq:total_theta}, which is given by the parameters of Eq.~\eqref{eq:d Lag} in the BMHV scheme,
\begin{align}
    &
        \theta_G (Q)
        +
        \delta \theta^{(1)}
        +
        \delta \theta^{(2)}_a
        +
        \delta \theta^{(2)}_b
        +
        \delta \theta^{(2)}_c
        +
        \delta \theta^{(2)}_{\rm c.t.}
        +
        \delta_{\theta_G}
        \nonumber \\
    &=
        \theta_G (Q)
        +
        \left[
            - 
            \frac{\opn{Im} [m_q]}{\opn{Re} [m_q]}
        +
            \im{\xi (Q)}
        \right]
        +
        \frac{\opn{Im} [y_q^2]}{16 \pi^2}
        (\opn{Re} [m_q])^2
        \left(
            I^{(2,2,1)}
            +
            2 \bar{I}^{(3,1,1)}
        \right)
        +
        \frac{\im{y_q^2}}{16 \pi^2}
        \frac{5}{8} \,.
\end{align}
The second and third terms in the right-handed side
%which is proportional to $\opn{Im} [y_q^2] (\opn{Re} [m_q])^2$ 
are the rephasing-invariant quantities, since they come from $\arg [m_q (1 + \xi)^*] \sim \frac{\im{m_q}}{\re{m_q}} - \im{\xi}$ and $\im{(y_q m_q^*)^2} \simeq \im{y_q^2} \re{m_q^2}$, respectively,  under the assumption $\re{m_q} \gg \im{m_q}$ and $|\xi| \ll 1$.
The remaining rephasing invariant is $\im{\left\{ y_q (1+\xi)^* \right\}^2}$.
Therefore, the last term would be rephasing-invariant, if we include terms proportional to $\im{y_q^2 \xi^*}$. They only arise effectively at the three-loop level and are beyond the scope of our current analysis.
The terms except for $\theta_G$ are rephasing invariants, so $\theta_G$ should also be, and it is unlike $\theta_G^{(\textrm{4-dim})}$ in Eq.~\eqref{eq:chiral anomaly}.
We guess that the $\theta_G$ is invariant in the BMHV scheme.
This is because the Jacobians for the integral measure which is source of the chiral anomaly are trivial under the local field redefinition in dimensional regularization. 
As also mentioned in Ref.~\cite{Valenti:2022uii}, the bare $\theta_G$ parameter ($\theta_G^0$) cannot be regarded as a dimensionless parameter of an axial transformation, due to the fact that it has mass dimension $-2 \epsilon$ (see Eq.~\eqref{eq:theta_G bare}) in dimensional regularization.
Indeed, the Fujikawa method relies on the heat-kernel regularization with the dimension set to $4$ strictly, not the dimensional one.
On the other hand, when we introduce the effective $\theta$ angle in Eq.~\eqref{eq:effective theta}, $\theta_G'^{(\rm 4-dim)}$, the full result Eq.~\eqref{eq:final_eq} can be understood by $4$-dimensional rephasing invariants.
$\theta_G'^{(\rm 4-dim)}$ is not a rephasing invariant, because its ingredient $\im{\xi}$ transforms under the chiral transformation.
Hence, $\theta_G'^{(\rm 4-dim)}$ has same transformation as $\theta_G^{(\rm 4-dim)}$, and the result is consistent with the previous work \cite{Banno:2023yrd}.

\begin{comment}
    
\tk{TK summary}

renormalization scale invariant combination: $\bar{\theta}^\prime_G$

renormalization scale dependent: $\bar{\theta}^\prime_G + \frac{\text{Im}y^2}{16 \pi^2} \frac{5}{8}$

chiral invariant combination: $\bar{\theta}^\prime_G - \arg [m_q]$, $\frac{\text{Im}y^2}{16 \pi^2} (\text{Re}m)^2$

chiral dependent(??): $\bar{\theta}^\prime_G - \arg [m_q] + \frac{\text{Im}y^2}{16 \pi^2} \frac{5}{8} \to \bar{\theta}^\prime_G - \arg [m_q] + \frac{\text{Im}(y (1+\xi^\ast))^2}{16 \pi^2} \frac{5}{8} $ (invariant?)

%how about the renormalization scale dependence of $ \arg [m_q]$? (I think under the QCD,  $ \arg [m_q]$ is renormalization scale-independent quantity, while it is scale dependent in our toy model.) 

So, 
physical observables: $\bar{\theta}^\prime_G - \arg [m_q] + \frac{\text{Im}(y (1+\xi^\ast))^2}{16 \pi^2} \frac{5}{8} + \frac{\text{Im}y^2}{16 \pi^2} (\text{Re}m)^2$, chiral invariant but scale dependent and also unpredictable (the value is fixed only by measurement)  (...like electron coupling constant, very reasonable!).
 
\end{comment}
\appendix
%=======================================================
%        ACKNOWLEDGEMENTS
%=======================================================
\acknowledgments
We are grateful for the fruitful discussions we had with Yukinari Sumino and Masaharu Tanabashi.
This work is supported by the JSPS Grant-in-Aid for Scientific Research Grant No.\,23K20232 (J.H.), No.\,24K07016 (J.H.), No.\,25H02180 (J.H.), No.\,24K22872 (T.K.), and No.\,25K07276 (T.K.). The work of J.H.\ is also supported by 
World Premier International Research Center Initiative (WPI Initiative), MEXT, Japan.
This work is also supported by 
JSPS Core-to-Core Program Grant No.\,JPJSCCA20200002. 
This work of N.O. was supported by JSPS KAKENHI Grant Number 24KJ1256.
This work was financially supported by JST SPRING, Grant Number JPMJSP2125.
The authors T.B. and K.O. would like to take this opportunity to thank the ``THERS Make New Standards Program for the Next Generation Researchers.''

%=======================================================
%        REFERENCES
%=======================================================
\bibliographystyle{utphys28mod}
\bibliography{ref}

\end{document}